\documentstyle[epsfig,multicol,aps,prl]{revtex} 
\renewcommand{\narrowtext}{\begin{multicols}{2}  
\global\columnwidth20.5pc}  
\renewcommand{\widetext}{\end{multicols}  
\global\columnwidth42.5pc}    
\newcommand{\pprl}{Phys. Rev. Lett. \ } 
\newcommand{\pprb}{Phys. Rev. {B}} 
\newcommand{\eq}{\begin{equation}}
\newcommand{\ee}{\end{equation}}
\newcommand{\eqa}{\begin{eqnarray}}
\newcommand{\eea}{\end{eqnarray}}

\newcommand{\tj}{{\rm t-J}}
\newcommand{\gbcs}{{\gamma{\rm BCS}}}
\newcommand{\bcs}{{{\rm BCS}}}

\renewcommand{\v}[1]{{\bf #1}}

\newcommand{\tp}{t^\prime}
\newcommand{\s}{\sigma}
\newcommand{\nocc}{n_{\rm occ}}
\begin{document}
\draft
\title{Local Moment Formation in the Superconducting 
State of a Doped Mott Insulator}

\author{Ziqiang Wang$^a$ and Patrick A. Lee$^b$}
\address{$^a$Department of Physics, Boston College, Chestnut Hill, MA 02467}
\address{$^b$Department of Physics, Massachusetts Institute of Technology,
Cambridge, MA 02139}

\date{\today}
\maketitle
\begin{abstract}
A microscopic theory is presented for the local moment formation
near a non-magnetic impurity or a copper defect in high-T$_c$ superconductors.
We use a renormalized meanfield theory of the $\tj$ model for a doped Mott 
insulator and study the fully self-consistent, spatially unrestricted 
solutions of the d-wave superconducting (SC) state in both the spin $S=0$ 
and $S=1/2$ sectors. We find a transition from the singlet d-wave SC state to
a spin doublet SC state when the renormalized exchange coupling 
exceeds a doping dependent critical value.
The induced $S=1/2$ moment is staggered and localized 
around the impurity. It arises from the binding of an 
$S=1/2$ nodal quasiparticle excitation to the impurity. 
The local density of states spectrum is calculated and connections to NMR and 
STM experiments are discussed.

\end{abstract}
\narrowtext

The local electronic structure near non-magnetic impurities in
high-$T_c$ superconductors has attracted wide attention recently.
A series of NMR experiments in YBa$_2$Cu$_3$O$_{7-x}$ (YBCO)
have shown that with the substitution of
nonmagnetic Zn/Li for the Cu atoms in the CuO$_2$ plane,
an $S=1/2$ staggered magnetic moment is generated on the Cu ions 
in the vicinity of the impurity sites
\cite{bobroff99,julien00,bobroff01,mahajan94-00,mendels99,mac00,fink90,kilian}.
Below the superconducting transition temperature,
the moment is partially screened in optimally doped YBCO, but
remains unscreened down to the lowest temperatures in the underdoped
regime \cite{bobroff01,julien00}.
Low-temperature STM experiments \cite{pan} have directly observed the 
local electronic structure around the Zn impurity atoms on the surface of 
superconducting Bi$_2$Sr$_2$CaCu$_2$O$_{8+x}$ (BSCCO). The tunneling
differential conductance above the Zn site exhibits a sharp in-gap resonance 
peak near zero bias (-1.5meV) for electrons tunneling out of the sample
\cite{pan}.

While the presence of a resonance impurity state 
is consistent with the strong local potential scattering 
near the impurity \cite{balatsky,flatte}, the latter does not
account for the local magnetic moment induced by nonmagnetic
impurities observed by NMR. This has led
to attempts to identify the low energy conductance peak with  
the Kondo resonance resulting from the Kondo screening of the local
moment by the superconducting (SC) electrons \cite{sachdev}. 
However, the {\it formation} of the local moment itself
near nonmagnetic impurities in a d-wave superconductor has not 
been addressed. In this paper, we present a theory which simultaneously
explains the local moment formation and predicts a strong low energy
resonance in the local density of states (LDOS) spectrum.

We study the local moment formation
in d-wave superconductors in proximity to Mott insulators. In the
resonance-valence-bond (RVB) picture \cite{anderson}, 
the electrons are paired into spin singlets, which are mobilized by the 
doped holes and condense into the SC state below $T_c$. 
A nonmagnetic impurity atom such as Zn/Li or a Cu vacancy locally breaks a 
singlet, creating an unpaired spin.
This extra $S=1/2$ moment can be self-consistently trapped by the local
impurity potential, forming a local moment confined to the
neighboring sites of the impurity. Our strategy is as follows.
The d-wave BCS ground state $\vert\bcs\rangle$ is a spin singlet and 
the low-lying excitations are the nodal quasiparticles. A single quasiparticle
excitation, $\gamma^\dagger\vert\bcs\rangle$, carries a spin $S=1/2$.
The unpaired spin induced by the impurity 
corresponds to a superposition of such one-$\gamma$ excitations. We
refer to such a doublet state as a $\gbcs$ state. 
If the self-consistently determined energy of the $\gbcs$ state
is lower than that of the $\bcs$ ground state,
the true ground state is a doublet with a net $S=1/2$ moment.
Below, we show that the physical picture
described above can be realized by a renormalized slave boson meanfield
theory of the $\tj$ model. We carry out fully self-consistent, spatially 
unrestricted calculations of the
valence bond, SC order parameter, local doping
concentration (LDC), and the spin density distribution in the doublet
$\gbcs$ state and compare its energy to that of the singlet d-wave BCS state.
We find that binding occurs
and the doublet $\gbcs$ state becomes the true ground state
when the renormalized exchange coupling $\alpha J$ exceeds a
nonzero doping dependent critical value.
The $S=1/2$ moment exhibits a staggered pattern localized around the 
impurity in agreement with the results derived from the NMR experiments. 
We determine the phase boundary in the $\alpha J$ versus doping plane,
obtain the tunneling density of states 
in the doublet $\gbcs$ state
and compare to the results of STM experiments.

We emphasize the importance for studying ground states with good 
spin quantum numbers. In a recent work \cite{Tsuchiura}, Gutzwiller 
approximation of the t-J model was used to calculate the induced moment 
near a nonmagnetic impurity. However, the authors did not seem to have
considered states with well defined spin quantum numbers. They reported
a net moment of a fraction of an electron spin which was obtained due to
a partial polarization of the singlet BCS state in the 
meanfield theory. Therefore, the binding of an spin-1/2 excitation 
to the impurity i.e. the formation of the local moment has not been addressed.

We begin with the $\tj$ model on a square lattice
in the presence of the long-range Coulomb interaction,
\eqa
H=&-&{\sum_{i,j}}^\prime (t_{ij}c_{i\s}^\dagger c_{j\s}+{\rm h.c.})
+J{\sum_{\langle i,j\rangle}}^\prime(\v S_i\cdot\v S_j\!-{1\over4}n_i n_j)
\nonumber \\
&+&{V_c\over2}{\sum_{i\ne j}}^\prime  {n_j-{\bar n}\over\vert
\v r_i-\v r_j\vert} n_i-\mu {\sum_i}^\prime n_i.
\label{htj}
\eea
Here $c_{i\s}^\dagger$ creates an electron and the spin operator 
$\v S_i=({1\over2})c_{i\alpha}^\dagger{\vec \sigma}_{\alpha\beta}c_{i\beta}$.
The sum over $i,j$ in the hopping term includes the nearest neighbors (n.n.)
where $t_{ij}\equiv t$ and the next n.n. where $t_{ij}\equiv\tp$.
To model a nonmagnetic impurity or a Cu defect, we remove a site
on the $L\times L$ square lattice as indicated by the primes on 
the summations in Eq.~(\ref{htj}).
The total number of sites is then $N_s=L\times L -1$.
The long-range Coulomb interaction strength $V_c$ in Eq.~(\ref{htj}) 
is given by $V_c=e^2/4\pi\epsilon a\sim t$ \cite{wangetal}.
Eq.~(\ref{htj}) is only physical in the limit of 
strong on-site Coulomb repulsion, i.e. under
the constraint of no double occupancy,
$n_i=c_{i\s}^\dagger c_{i\s}\le1$. Such a projected
Hilbert space can be treated in the slave-boson formulation by
writing $c_{i\s}^\dagger=f_{i\s}^\dagger b_i$, where $f_{i\s}^\dagger$
is a spin-carrying fermion and $b_i$ a spinless boson
\cite{kotliarliu} with the constraints
$f_{i\s}^\dagger f_{i\s}+b_i^\dagger b_i=1$.
The AF spin-exchange
term is decoupled according to\cite{brinkmanlee}
\eqa
{\v S_i\cdot\v S_j} = &-& \frac{3}{8}[\chi_{ij}^*f_{i\s}^\dagger f_{j\s}
+ \Delta_{ij}^*(f_{i \downarrow} f_{j \uparrow}
-f_{i \uparrow}f_{j \downarrow})
+{\rm h.c.}]
\nonumber \cr
&+& \frac{3}{8}(\vert\chi_{ij}\vert^2+\vert\Delta_{ij}\vert^2),
\label{ss}
\eea
where $\Delta_{ij}$ and $\chi_{ij}$ are the pairing and valence bond
order parameters on each n.n. bond respectively,
\eq
\Delta_{ij}=\langle f_{i \downarrow} f_{j \uparrow}
-f_{i \uparrow}f_{j \downarrow}\rangle,\quad
\chi_{ij} = \langle f_{i\s}^\dagger f_{j\s}\rangle.
\label{deltachi}
\ee
In the meanfield SC state,
the bosons locally condense \cite{wangetal,kotliarliu} into the coherent state,
\eq
\vert\Psi_b\rangle=A e^{\sum_i^\prime\phi_i b_i^\dagger}
\vert0\rangle, \qquad A=e^{-{1\over2}\sum_i^\prime\vert\phi_i\vert^2}.
\label{psib}
\ee
The LDC is given by $x_i=\langle\Psi_b\vert b_i^\dagger b_i\vert\Psi_b\rangle
=\vert\phi_i\vert^2$. The constraint $\vert\phi_i\vert^2=1-n_i^f$ 
is enforced on average at every site by locally
adjusting the chemical potential.

It is known that fluctuations beyond such a meanfield RVB state
will mediate a residual direct spin-spin interaction,
which can be approximately accounted for phenomenalogically
by adding to the meanfield theory an interaction term,
\eq
H_\alpha=\alpha J {\sum_{\langle i,j\rangle}}^\prime
\langle \v S_i\rangle \cdot\v S_j,
\label{halpha}
\ee
where $\alpha$ is a phenomenalogical parameter between zero and one.
This renormalized meanfield approach was used recently for
a successful description of the spin dynamics in YBCO \cite{brinkmanlee}.
The fully self-consistent, spatially unrestricted
solution will include nonlinear magnetic effects induced by $H_\alpha$ 
near the impurity.

We diagonalize the theory in real space by writing
\eq
\{f_{i\uparrow}^\dagger(t),f_{i\downarrow}(t)\}
=\sum_n \{u_{ni},v_{ni}\}\gamma_n^\dagger e^{-iE_nt/\hbar},
\label{uv}
\ee
where $\gamma_n^\dagger$, $n=1,\dots,2N_s$ creates a
Bogoliubov quasiparticle with energy $E_n$ and
wave-function $(u_{ni},v_{ni})$ as solutions of
the Bogoliubov-de Gennes (BdG) equations.
Since the particle-hole symmetry in the quasiparticle excitation
spectrum, which reflects the degeneracy in spin excitations,
is broken in the presence of a net spin moment,
we keep all independent eigenstates
with $E_n<0$ and $E_n>0$ and construct the ground state according to
the occupation of the fermionic quasiparticle orbitals:
\eq
\vert \Psi_G, \nocc\rangle \! =\! \prod_{n=1}^{\nocc}\gamma_n^\dagger\vert 
0\rangle
\!=\!\prod_n^{\nocc}\sum_{i=1}^{N_s}[u_{ni}^* f_{i\uparrow}^\dagger
+ v_{ni}^* f_{i\downarrow}]\vert 0\rangle,
\label{gs}
\ee  
where $\nocc$ labels the last $\gamma$-quasiparticle level to be occupied.
The spin-singlet d-wave BCS state corresponds to
$\vert \bcs \rangle=\vert \Psi_G, \nocc=N_s\rangle$, whereas the $\gbcs$ state
is given by $\vert \gbcs \rangle=\vert \Psi_G, \nocc=N_s+1\rangle$, which is
a global doublet with a net $S=1/2$ moment. 
These states must be
determined by minimizing $\langle\Psi_G,\nocc\vert H_{\rm mf}+H_\alpha
\vert\Psi_G,\nocc\rangle$, leading to the
self-consistency equations
\eqa
\Delta_{ij}=\sum_{n=\nocc+1}^{2N_s} v_{ni}^* u_{nj}
-\sum_{n=1}^{\nocc}u_{ni} v_{nj}^*,
\label{dijuv} \\
\chi_{ij}=\sum_{n=\nocc+1}^{2N_s} v_{ni}^*v_{nj}
+\sum_{n=1}^{\nocc} u_{ni}u_{nj}^*.
\label{chiijuv}
\eea
The LDC and the local spin density are given by
\eqa
x_i=
1-\sum_{n=\nocc+1}^{2N_s}\vert v_{ni}\vert^2 
-\sum_{n=1}^{\nocc} \vert u_{ni}\vert^2,
\label{holedensity} \\
S_i^z={1\over2}\left[\sum_{n=1}^{\nocc} \vert u_{ni}\vert^2
-\sum_{n=\nocc+1}^{2N_s}\vert v_{ni}\vert^2\right].
\label{spindensity}
\eea
For $ \vert \Psi_G, \nocc\rangle$ to be a self-consistent
ground state, it requires a lowest total energy and, in addition,
$E_{\nocc}<0$.  

The BdG equation and the spatially unrestricted
self-consistency equations can be solved by numerical iterations starting
from a random set of initial conditions.
We use $J$ as the unit of energy and set $t/J=3$, $V_c/J=5$.
The impurity/defect is located at $(L/2,L/2)$. We first consider $\tp=0$ and
focus on the nature of the ground state. The energies per site
for the $\bcs$ and the $\gbcs$ states are shown 
in Fig.~1 as a function of the renormalized spin-spin interaction 
$\alpha J$ for $L=24$ at an average doping $x=0.14$.
While that of the singlet state is independent of $\alpha J$, the
energy of the doublet $\gbcs$-state is reduced with increasing $\alpha J$
and drops abruptly below that of the singlet state near a critical value 
$\alpha_c$, indicating that the doublet state can gain substantial
spin-exchange energy through the formation of the magnetic moment.
Notice that although the doublet state energy is already lower
than that of the singlet state for $\alpha$ less than $\alpha_c$,
it is not a self-consistent ground state unless the last occupied
$N_s+1$ quasiparticle level is pulled below the Fermi level 
(see Eq.~(\ref{gs})).
To determine the value of $\alpha_c$ at which a phase transition
into the $\gbcs$-state occurs, we calculate the energy
of the $\nocc=N_s+1$-th $\gamma$-quasiparticle level $E_{\nocc}$
as a function of $\alpha J$ and identify $\alpha_c\simeq0.68$ at $x=0.14$
by the point where it crosses the Fermi energy.
The phase diagram in the plane of $\alpha J$ versus doping $x$
is plotted in the inset of Fig.~1, which shows that
the formation of the local moment requires a smaller
renormalized spin interaction in the underdoped regime. 
\begin{figure}      
\vspace{-0.5truecm}  
\center      
\centerline{\epsfysize=2.8in      
\epsfbox{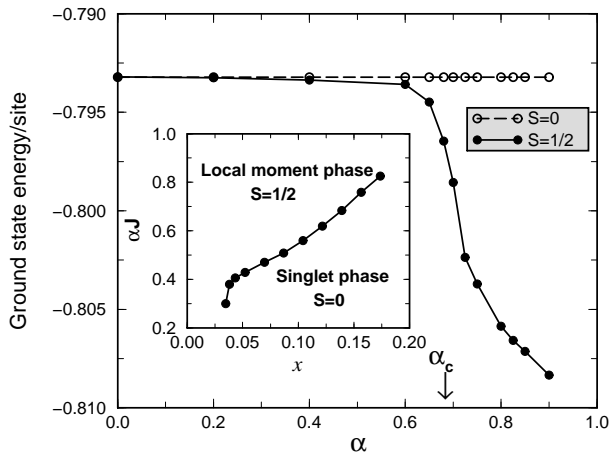}}  
\vspace{-0.5truecm}      
\begin{minipage}[t]{8.1cm}      
\caption{Energy per site of the singlet and the doublet
states as a function of $\alpha J$ on a $24\times24$ lattice at
$x=0.14$. The critical value of the exchange coupling
is $\alpha_c\simeq0.68$.
Inset: phase diagram in the $\alpha J$ versus doping $x$ plane.}
\label{fig1}     
\end{minipage}      
\end{figure}      

Next, we turn to the spatial distribution of the impurity/defect induced
local moment in the $\gbcs$-state. We choose large systems with $L=32$ 
and set $\tp=-t/3$ and $\alpha=0.75>\alpha_c$ at $x=0.14$. 
The self-consistent solutions of the LDC
$x_i$ and the d-wave gap amplitude, $\Delta_i^d
=(\Delta_{i,i+{\hat x}}+\Delta_{i,i-{\hat x}}
-\Delta_{i,i+{\hat y}}-\Delta_{i,i-{\hat y}})/4$,
are shown in Figs.~2a and 2b as 2D maps. Both $x_i$ and $\Delta_i^d$
are strongly suppressed in the vicinity of the impurity/defect, but
quickly resume their bulk values away from it. 
The valence bond $\chi_{ij}$ shows only moderate oscillations near
the impurity with no signatures for the emergence of a staggered 
flux or current. The spatial distribution
of the magnetic moment $S_i^z$, having a net magnitude $\sum_i^\prime
S_i^z=1/2$, is plotted in Fig.~2c. It shows a staggered pattern
localized around the impurity in good agreement with that derived from
the NMR experiments \cite{julien00,bobroff01}. The majority of the net
moment resides on the four nearest neighbors of
the impurity. Interestingly, along the 
nodal directions (diagonals) of the d-wave gap,
weak incommensurate spin density wave (SDW) oscillations arise
with a periodicity of approximately $8$ unit cells.
This feature is more clearly seen in the 3D plot of the {\it staggered}
magnetic moment shown in Fig.~2d. The dependence of the
SDW periodicity on the value of $\tp$ and the doping $x$
is consistent with the behavior of the
bulk spin susceptibility which peaks around similar incommensurate
wave-vectors \cite{brinkmanlee}. Physically, an impurity/defect
induces a local moment in its immediate vicinity, which subsequently
leaks into the bulk via the short-range spin-spin correlation in the
short-range RVB state.
\begin{figure}      
\vspace{-0.5truecm}  
\center      
\centerline{\epsfysize=2.8in \epsfbox{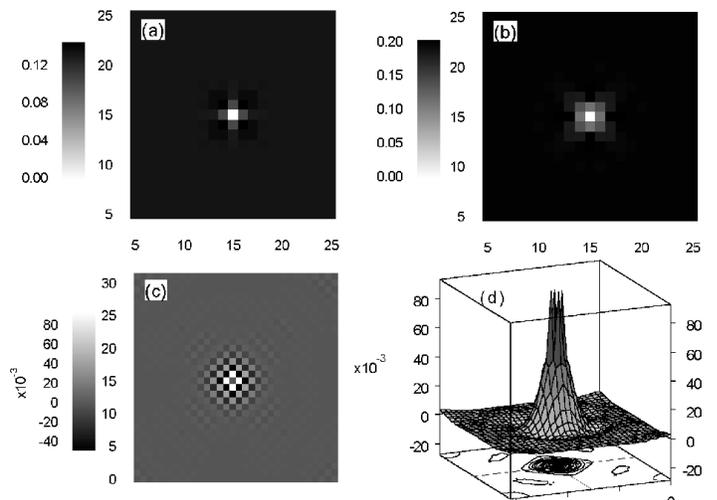}}
\vspace{-0.5truecm}      
\begin{minipage}[t]{8.1cm}      
\caption{Self-consistent solution of the $\gbcs$-state with
a net $S=1/2$ local moment on a $32\times32$ lattice
with $\alpha=0.75>\alpha_c$ and $\tp=-t/3$. 
The impurity/defect is located at $(15,15)$.
(a) LDC $x_i$. (b) d-wave gap amplitude
$\Delta_i^d$. (c) Distribution of the local moment $S_i^z$.
(d) 3D-plot of the {\it staggered} magnetic moment.
}    
\label{fig2}     
\end{minipage}      
\end{figure}      

We next study the LDOS spectrum measured by STM.
Note that the net spin polarization in the doublet ground state flips
between $S^z=\pm1/2$, which should be identified with the two degenerate 
states of an $S=1/2$ quantum impurity. We set $S^z=1/2$ and do not consider 
here the effect of Kondo coupling between such an extended local moment and the
quasiparticles which can be quite different from the conventional 
Kondo problem in d-wave superconductors. 
The spectral weight of the electron is a product of
those associated with the boson and fermion states given in
Eqs.~(\ref{psib}) and (\ref{uv}) respectively. We derive
the LDOS for the up and down spin electrons,
\eqa
N_{i\uparrow}(\omega)&=&\bigl[\sum_n^{\nocc}A_{n\uparrow}^- \vert u_{ni}\vert^2
+\sum_{n={\nocc}+1}^{2N_s}A_{n\uparrow}^+\vert u_{ni}\vert^2
\bigr]\delta(\omega-E_n)
\nonumber \\
N_{i\downarrow}(\omega)&=&\bigl[
\sum_n^{\nocc}A_{n\downarrow}^+\vert v_{ni}\vert^2
+\sum_{n={\nocc}+1}^{2N_s}A_{n\downarrow}^-\vert v_{ni}\vert^2\bigr]
\delta(\omega+E_n),
\nonumber
\eea
where $A_{n\sigma}^{\pm}$ is the boson coherent state overlap
for adding ($+$) or removing ($-$) an electron with spin-$\sigma$
accompanied by adding/removing a $\gamma$-quasiparticle from the
$n$-th fermionic orbital. In order to maintain the occupation
constraint in the final states, the corresponding boson coherent states,
defined as $\phi_{in\sigma}^{\pm}$, must satisfy
$(\phi_{in\uparrow}^{\pm})^2=(\phi_{in\downarrow}^{\mp})^2
=\phi_{i}^2\pm\vert
u_{ni}\vert^2\mp\vert v_{ni}\vert^2$.
Using Eq.~(\ref{psib}), we obtain the spectral weight for the bosons,
\eqa
A_{n\uparrow(\downarrow)}^+&=&\phi_{i}^2e^{-2\sum_i^\prime[\phi_{i}
-\phi_{in\uparrow(\downarrow)}^+]^2},
\nonumber \\
A_{n\uparrow(\downarrow)}^-&=&[\phi_{in\uparrow(\downarrow)}^-]^2
e^{-2\sum_i^\prime[\phi_{i}-\phi_{in\uparrow(\downarrow)}^-]^2}.
\nonumber
\eea
\begin{figure}      
\vspace{-0.5truecm}  
\center      
\centerline{\epsfysize=2.8in \epsfbox{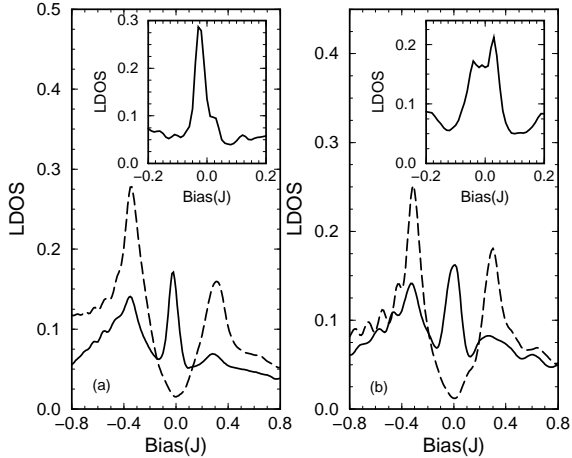}}
\vspace{-0.5truecm}      
\begin{minipage}[t]{8.1cm}      
\caption{ 
The LDOS spectra at the n.n.
sites of the impurity/defect (solid lines) and near the corner 
(dashed lines) of $32\times32$ lattices at $x=0.14$.
(a) $\tp=-0.45t$ and $\alpha J=0.7$, close to the phase boundary.
(b) $\tp=-t/3$ and $\alpha J=0.75$, the same system as in Fig.~2.
Insets: details of the resonance line-shape near zero bias with
reduced thermal broadening.
}    
\label{fig3}     
\end{minipage}      
\end{figure}      
Due to the localized nature of the
impurity states, the order parameters in
Eqs.~(\ref{dijuv}-\ref{spindensity}) are insensitive to the
boundary conditions for large system sizes.
We have calculated $N_i(\omega)$ on $32\times32$ lattices
and averaged the results over a set of different boundary conditions 
to reduce the finite-size effects. The complex boundary phases along
the $x$ and $y$ directions are typically $2\pi(i_x/5,i_y/5)$ where
$i_x,i_y=0,\dots,4$. The results are not sensitive to the detailed
choice of the phases. In Fig.~3, we plot the LDOS as a function
of the bias voltage for two ground states, one is close to the phase boundary 
(Fig.~3a) and the other (Fig.~3b) is the same system used in Fig.~2 
in the local moment phase. On the n.n. site of the impurity/defect, 
the LDOS (solid lines) exhibits a near zero-bias resonance with its peak 
located on the occupied side of the spectrum.
Comparing to the LDOS obtained near the corner of the system 
(dashed lines), the
coherence peaks at the gap edges have been strongly 
suppressed around the impurity/defect. These results are 
in good agreement with STM observations \cite{pan}.
The LDOS spectra with reduced thermal broadening 
are plotted in the insets of Fig.~3 to reveal the details of the
resonance line-shape. The inset in Fig.~3a is in good agreement with current
experiments \cite{pan}. However, comparison with the inset in Fig.~3b 
shows that the resonance line-shape is sensitive to the parameters and show
several peaks asymmetrically
distributed around zero bias. The low energy spectra  can be qualitatively 
understood by considering the last two occupied quasiparticle levels: 
$E_{N_s}$ and $E_{N_s+1}$. 
The resonance state at these two energies is occupied
in the local moment phase which should in principle lead to
four resonance peaks centered at $\pm E_{N_s}$ and
$\pm E_{N_s+1}$ in the LDOS spectrum through their particle-hole
reflections. However, since the SC coherence is
significantly reduced near the impurity/defect, the spectral weights
have the property $\vert u_{N_s}\vert^2\ll \vert v_{N_s}\vert^2$
whereas $\vert u_{N_s+1}\vert^2\gg\vert v_{N_s+1}\vert^2$
leading to two predominant peaks near zero bias.
The peak at the negative energy $E_{N_s+1}$ (positive energy 
$\vert E_{N_s}\vert$) corresponds to
removing (adding) an up (down) spin electron from the resonance state. 

In conclusion, we have presented a microscopic theory that describes
on equal footing the formation of local moment observed by NMR 
and the local electronic structure measured by STM 
near nonmagnetic impurities in high-T$_c$ superconductors.
The existence of the low energy resonance
is a robust feature of our model, even though the detailed structure
of the resonance may vary. Unless the {\it surface} of BSCCO is very
close to the onset of local moment formation such that the detailed
resonance peak structure becomes difficult to observe,
our model predicts multiple peak structures for
better resolved resonances in the local moment
phase. Moreover, since recent STM revealed a
spatially inhomogeneous electronic structure in BSCCO \cite{pan2}
that can be interpreted as off-stoichiometry carrier doping induced 
spatial variations in the LDC \cite{wangetal},
the present theory would predict a distribution of 
the resonance structure over different Zn-impurity sites.

We thank X. Dai, E. Hudson, and S.-H. Pan for useful discussions. 
P.A.L is supported by NSF Grant No. DMR-0201069.
Z.W. is supported by DOE Grant No. DE-FG02-99ER45747 and an award 
from Research Corporation.

\end{multicols} 
\end{document}